\documentclass[12pt]{emulateapj}
\usepackage{natbib}

\newcommand{\etal}{et~al.}
\newcommand{\eg}{e.g.,}

\newcommand{\spitzer}{{\it Spitzer}}
\newcommand{\chandra}{{\it Chandra}}
\newcommand{\hst}{{\it HST}}

\newcommand{\msun}{$M_\odot$}
\newcommand{\mathmsun}{M_\odot}

\newcommand{\clA}{ISCS~J1438.1+3414}
\newcommand{\clB}{ISCS~J1432.4+3250}
\newcommand{\clname}{SPT-CL~J0546-5345}

\newcommand{\rtwo}{r_{\mbox{\scriptsize 200}}}

\newcommand{\Mfive}{M_{\mbox{\scriptsize 500}}}
\newcommand{\Mtwo}{M_{\mbox{\scriptsize 200}}}

\newcommand{\Mtwolx}{$M_{\mbox{\scriptsize 200,$L_X$}}$}
\newcommand{\Mfivelx}{$M_{\mbox{\scriptsize 500,$L_X$}}$}

\newcommand{\Mtwodyn}{M_{\mbox{\scriptsize 200,dyn}}}
\newcommand{\Mtwowl}{M_{\mbox{\scriptsize 200,WL}}}

\newcommand{\mathMtwolx}{M_{\mbox{\scriptsize 200,$L_X$}}}
\newcommand{\mathMtwotx}{M_{\mbox{\scriptsize 200,$T_X$}}}
\newcommand{\mathMfivelx}{M_{\mbox{\scriptsize 500,$L_X$}}}
\newcommand{\logmlxtwo}{\log{(\mathMtwolx/\mathmsun)}}
\newcommand{\logmtxtwo}{\log{(\mathMtwotx/\mathmsun)}}
\newcommand{\logmlxfive}{\log{(\mathMfivelx/\mathmsun)}}
\newcommand{\logmtwo}{\log{(\Mtwo/\mathmsun)}}
\newcommand{\logmdyntwo}{\log{(\Mtwodyn/\mathmsun)}}
\newcommand{\logmwltwo}{\log{(\Mtwowl/\mathmsun)}}

\def\spose#1{\hbox to 0pt{#1\hss}}
\def\simlt{\mathrel{\spose{\lower 3pt\hbox{$\mathchar"218$}}
     \raise 2.0pt\hbox{$\mathchar"13C$}}}
\def\simgt{\mathrel{\spose{\lower 3pt\hbox{$\mathchar"218$}}
     \raise 2.0pt\hbox{$\mathchar"13E$}}}

\slugcomment{Submitted to ApJ}

\shorttitle{X-Ray Emission from Two ISCS Clusters at $z>1.4$}
\shortauthors{Brodwin et al.}

\newcommand{\CfA}{1}
\newcommand{\KeckFellow}{2}
\newcommand{\JPL}{3}
\newcommand{\Davis}{4}
\newcommand{\LLNL}{5}
\newcommand{\UFlorida}{6}
\newcommand{\MIT}{7}
\newcommand{\NOAO}{8}
\newcommand{\Durham}{9}
\newcommand{\Harvard}{10}

\begin{document}


\title{X-Ray Emission from Two Infrared-Selected Galaxy Clusters at
  $z>1.4$ in the IRAC Shallow Cluster Survey}


\author{M.~Brodwin,\altaffilmark{\CfA, \KeckFellow}
D.~Stern,\altaffilmark{\JPL}
 A. Vikhlinin,\altaffilmark{\CfA}
 S.~A.~Stanford,\altaffilmark{\Davis,\LLNL}
 A.~H.~Gonzalez,\altaffilmark{\UFlorida}
 P.~R.~Eisenhardt,\altaffilmark{\JPL}
 M.~L.~N.Ashby,\altaffilmark{\CfA}  
 M.~Bautz,\altaffilmark{\MIT} 
 A.~Dey,\altaffilmark{\NOAO} 
 W.~R.~Forman,\altaffilmark{\CfA} 
 D.~Gettings,\altaffilmark{\UFlorida} 
 R.~C.~Hickox,\altaffilmark{\Durham}
 B.~T.~Jannuzi,\altaffilmark{\NOAO}
 C.~Jones,\altaffilmark{\CfA}
 C.~Mancone,\altaffilmark{\UFlorida}
 E.~D.~Miller,\altaffilmark{\MIT}
 L.~A.~Moustakas,\altaffilmark{\JPL}
 J.~Ruel,\altaffilmark{\Harvard}
 G.~Snyder,\altaffilmark{\CfA}
and G.~Zeimann\altaffilmark{\Davis}
}


\altaffiltext{\CfA}{Harvard-Smithsonian Center for Astrophysics, 60 Garden Street, Cambridge, MA 02138}
\altaffiltext{\KeckFellow}{W. M. Keck Postdoctoral Fellow at the Harvard-Smithsonian Center for Astrophysics}
\altaffiltext{\JPL}{Jet Propulsion Laboratory, California Institute of Technology, Pasadena, CA 91109}
\altaffiltext{\Davis}{Department of Physics, University of California, One Shields Avenue, Davis, CA 95616}
\altaffiltext{\LLNL}{Institute of Geophysics and Planetary Physics, Lawrence Livermore National Laboratory, Livermore, CA 94550}
\altaffiltext{\UFlorida}{Department of Astronomy, University of Florida, Gainesville, FL 32611}
\altaffiltext{\MIT}{Kavli Institute for Astrophysics and Space Research, MIT, Cambridge, MA 02139}
\altaffiltext{\NOAO}{NOAO, 950 North Cherry Avenue, Tucson, AZ 85719}
\altaffiltext{\Durham}{Department of Physics, Durham University, South~Road, Durham, DH1 3LE, UK}
\altaffiltext{\Harvard}{Department of Physics, Harvard University, 17 Oxford Street, Cambridge, MA 02138}


\begin{abstract}

  We report the X-ray detection of two $z>1.4$ infrared-selected
  galaxy clusters from the IRAC Shallow Cluster Survey (ISCS).  We
  present new data from the {\it Hubble Space Telescope} and the
  W.~M.~Keck Observatory that spectroscopically confirm cluster \clB\
  at $z=1.49$, the most distant of 18 confirmed $z>1$ clusters in the
  ISCS to date.  We also present new spectroscopy for \clA, previously
  reported at $z = 1.41$, and measure its dynamical mass.  Clusters
  \clB\ and \clA\ are detected in 36ks and 143ks \chandra\ exposures
  at significances of $5.2\sigma$ and $9.7\sigma$, from which we
  measure total masses of $\logmlxtwo = 14.4 \pm 0.2$ and
  $14.35\,^{+0.14}_{-0.11}$, respectively.  The consistency of the
  X-ray and dynamical properties of these high redshift clusters
  further demonstrates that the ISCS is robustly detecting massive
  clusters to at least $z = 1.5$.

\end{abstract}





\section{Introduction}

Present-day galaxy clusters contain large, old, roughly coeval
populations of massive, quiescent galaxies spanning a vast range of
local densities.  As such, they provide a natural laboratory in which
to test models of galaxy formation and evolution.  To trace the
evolution of cluster galaxies over their full lifetime, it is
necessary to identify and study the {\it precursor} cluster population
over a large redshift range.  For instance, the Coma cluster, with a
present-day mass of $\logmtwo \approx 15.3$ \citep{kubo07}, is
expected to have a precursor at $z \sim 1.5$ with a halo mass of
$\logmtwo \sim 14.6$.  This kind of archaeology requires uniformly
selected, well characterized cluster samples in which the evolutionary
precursors can be statistically identified, and which are sensitive
down to the group scale at very high redshift.  Neither X-ray nor
Sunyaev-Zel'dovich (SZ) cluster surveys have the mass sensitivity at
high redshift, and optical methods fail at $z\ga 1$ as the red
sequence shifts to the infrared.

The \spitzer/IRAC Shallow Cluster Survey \citep[ISCS,][]{eisenhardt08}
is a stellar mass-selected galaxy cluster survey spanning $0.1 < z <
2$. Clusters are identified via stellar mass overdensities in a 4.5
$\mu$m-selected galaxy sample using accurate photometric redshifts
\citep{brodwin06_iss}, and their selection is therefore independent of
the presence of a red sequence. There are 335 clusters and groups in
the sample, identified over 7.25 deg$^2$ within the Bo\"otes field of
the NOAO Deep Wide--Field Survey \citep{ndwfs99}, and 1/3 of the
clusters are at $z > 1$.  The ISCS cluster sample has a mean halo
mass, derived from its clustering, of $\approx 10^{14}$ \msun\ out to
$z=1.5$ \citep{brodwin07}.  Cluster photometric redshift accuracy,
based on comparison with over 100 clusters spanning $0< z< 1.5$, is
excellent, with $\sigma = 0.028(1 + z)$. At $z < 1$ roughly 100
clusters have been spectroscopically confirmed, primarily using the
extensive spectroscopic database of the AGN and Galaxy Evolution
Survey (AGES, Kochanek \etal\ in prep).  To date, we also have
spectroscopically confirmed 18 clusters spanning $1 < z < 1.5$.
(\citealt{stanford05,brodwin06_iss,elston06,eisenhardt08}, Brodwin
\etal\ in prep; Stanford \etal\ in prep). {\it All} of the candidates
for which adequate spectroscopy has been obtained have turned out to
be at the predicted photometric redshifts. The ISCS is therefore the
largest sample of spectroscopically confirmed galaxy clusters at $z >
1$.

In this paper, we report the X-ray detection of two of the most
distant ISCS clusters: \clA\ at $z=1.414$, first reported in
\citet{stanford05}, and \clB\ at $z=1.487$, for which we present
spectroscopic confirmation in this work.  The latter is the most
distant cluster to date to be spectroscopically confirmed in the ISCS.
We present the spectroscopic observations and dynamical properties of
these clusters in \textsection{\ref{Sec: dynamical}}.  In
\textsection{\ref{Sec: x-ray}} we present the X-ray observations, from
which we estimate total cluster masses.  In \textsection{\ref{Sec:
    discussion}} we compare the dynamical and X-ray properties of
these clusters, including testing the well-known relation between
velocity dispersion and temperature at the highest redshift yet.  We
also discuss the effect of possible systematic uncertainties on our
measurements.  We present our conclusions in \textsection{\ref{Sec:
    conclusions}}.  We use Vega magnitudes and a WMAP7+BAO+$H_0$
$\Lambda$CDM cosmology \citep{komatsu10}: $\Omega_M = 0.272$,
$\Omega_\Lambda = 0.728$, and $H_0 = 70.2$ km s$^{-1}$ Mpc$^{-1}$.

\section{Spectroscopy and Dynamical Masses}
\label{Sec: dynamical}

\subsection{\clA}
\begin{figure*}[bthp]
\epsscale{1.15}
\plotone{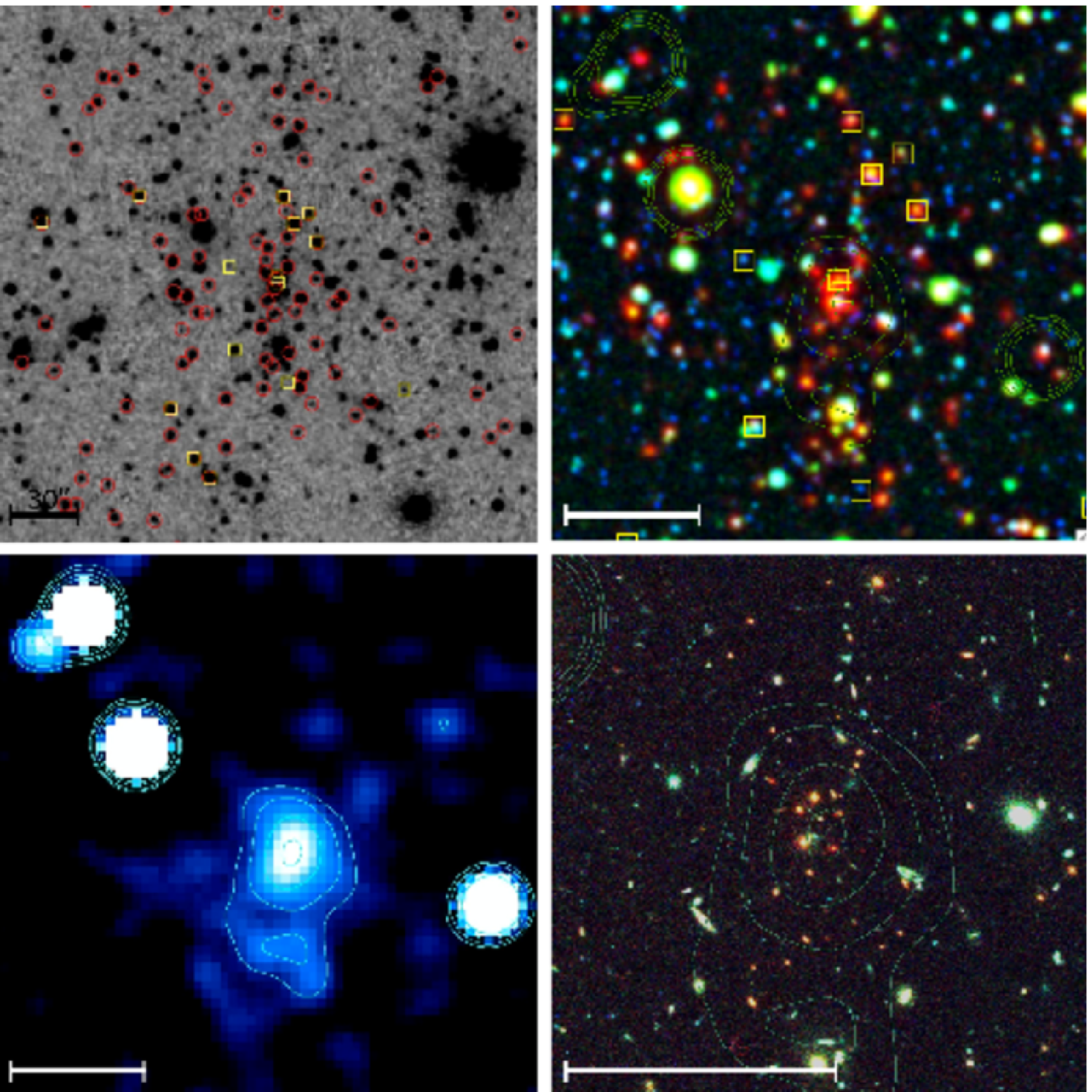}
\caption{{\small \spitzer, \chandra, and \hst\ and KPNO images of
    \clA\ at $z=1.414$. {\it Upper left panel:} $4\arcmin \times
    4\arcmin$ ($\approx 2 \times 2$ Mpc) IRAC 4.5$\mu$m image from the
    \spitzer\ Deep, Wide-Field Survey \citep[SDWFS;][]{ashby09}, with
    photometric (red circles) and spectroscopic (yellow boxes)
    redshifts indicated.  A galaxy is deemed a photometric redshift
    member if the integral of its redshift probability function about
    the cluster redshift exceeds 0.3 (see \citealt{eisenhardt08} for
    details). The overdensity in stellar mass at $z\sim 1.4$ is
    apparent.  {\it Upper right panel:} Pseudo-color optical ($B_WI$)
    + IRAC (4.5$\mu$m) image of the central $2\arcmin$ ($\approx 1$
    Mpc).  The contours are from the X-ray image shown in the panel at
    lower left.  The peak of the X-ray emission and the IRAC wavelet
    centroid are within 5\arcsec\ of each other.  {\it Lower left
      panel:} Binned (4$\times$4) \chandra\ image in the 0.7-2.0 keV
    energy band of the central $\approx 1$ Mpc region of \clA. The
    contours correspond to 0.60, 0.75, 1.05 and 1.40 counts per
    $2\arcsec \times 2\arcsec$ pixel in this energy range in 144.2 ks.
    {\it Lower right panel:} The cluster core, $1\arcmin$ ($\approx
    0.5$ Mpc) across centered on the X-ray centroid, shown in high
    resolution \hst\ images (ACS/F775W, ACS/F850LP, and WFC3/F160W).
    The X-ray contours lie on top of the red, early type galaxies in
    the core of \clA.  A 30\arcsec\ scale bar is given in each
    panel.}}
  \label{Fig: image22}
\end{figure*}

At the time of its discovery, cluster \clA\ at $z=1.414$
(Fig.~\ref{Fig: image22}), was the most distant galaxy cluster known
\citep{stanford05}.  To place this cluster in a proper evolutionary
context, it is imperative to directly measure its total mass.
\citet{eisenhardt08} previously reported the large stellar mass
content in \clA, inferred from the infrared luminosity of its members.
Recently, \citet{barbary10} published the SN Ia rate from a sample of
25 high redshift clusters selected by X-ray, optical and infrared
methods.  As the SN Ia rate is proportional to the cluster stellar
mass content, these clusters were selected to be the most massive
known at high redshift.  \clA\ is the only cluster in that work found
to contain {\it two} confirmed SNe Ia.  These SNe Ia, both hosted by
early-type galaxies, are the highest redshift cluster SNe known.
These results strongly suggest that \clA\ is not a group but rather a
bona fide massive cluster, with $\Mtwo > 10^{14}$ \msun.

\begin{figure}[bthp]
\epsscale{1.15}
\plotone{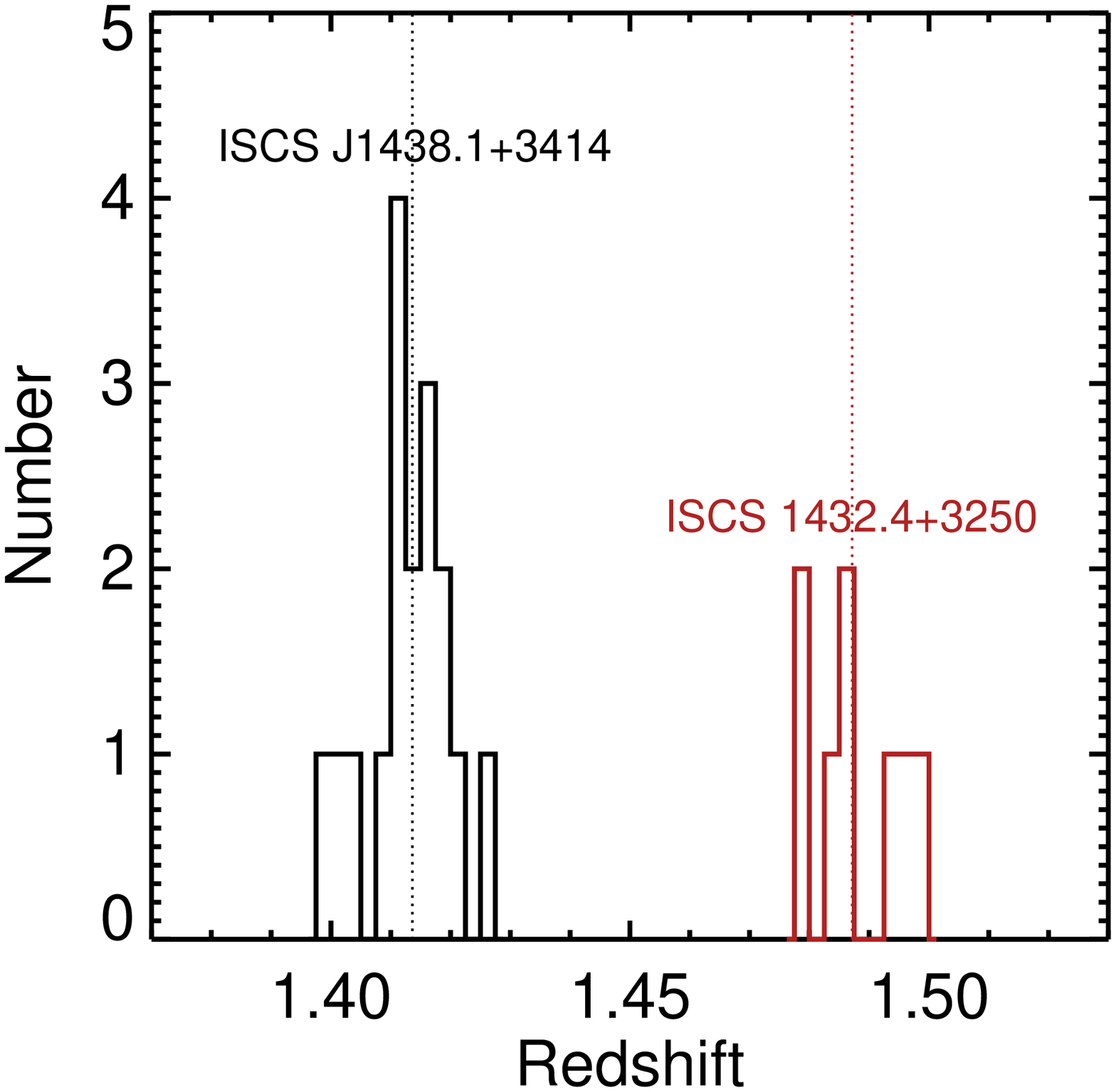}
\caption{Spectroscopic redshift histogram of confirmed members in
  clusters \clA\ (black) and \clB\ (red).  The well-sampled redshift
  distribution in \clA\ appears to trace a single massive halo with
  Gaussian-distributed velocities.}
  \label{Fig: zhist}
\end{figure}

Using a combination of optical multi-object Keck spectroscopy and
infrared \hst/WFC3 grism spectroscopy, we have secured 17 confirmed
cluster members in \clA\ within a radius of 2\,Mpc, listed in Table
\ref{Tab: spectra}.  Spectra with the Low-Resolution Imaging
Spectrograph (LRIS; \citealt{lris}) on Keck I were acquired on UT 2005
June 4 using $1.3\arcsec \times 10\arcsec$ slitlets, the G400/8500
grating on the red side, the D560 dichroic, and the 400/3400 grism on
the blue side.  The seeing was 0\farcs9 and conditions were clear.
The Deep Imaging Multi-Object Spectrograph (DEIMOS; \citealt{deimos})
spectra, acquired on UT 5 April 2007, used the 600ZD grating, the
OG550 order sorting filter and 1\arcsec\ slitlets.  A complete
description of the DEIMOS observations is given in \citet{dawson09}.
The infrared \hst\ spectroscopy was obtained on UT 2010 May 22 with
the Wide Field Camera 3 (WFC3) using the G141 grism and reduced using
the aXe software package \citep{aXe}.  Further details on the
spectroscopic observations for both of these clusters will be
presented in forthcoming papers.

The redshift histogram of cluster members is shown in Figure \ref{Fig:
  zhist}.  From these we calculate a biweight \citep{beers90} velocity
dispersion of $\sigma = 757\,^{+247}_{-208}\,$ km s$^{-1}$, including
the relativistic correction and the usual correction for velocity
errors \citep{danese80}.  The uncertainty estimates, obtained from
bootstrap resampling, represent the 68\% confidence interval.  Both
the gapper \citep{beers90} and simple standard deviation estimates
yield essentially identical dispersions.  Under the assumption that
\clA\ is approximately virialized, we use the mass-dispersion relation
from \citet{evrard08} to infer a dynamical mass of $\logmdyntwo =
14.4\,^{+0.3}_{-0.7}$.  This is in excellent agreement with the weak
lensing mass, $\logmwltwo = 14.5 \pm 0.3$ (Jee, priv.~comm).

\subsection{\clB}

\begin{figure*}[bthp]
\epsscale{1.15}
\plotone{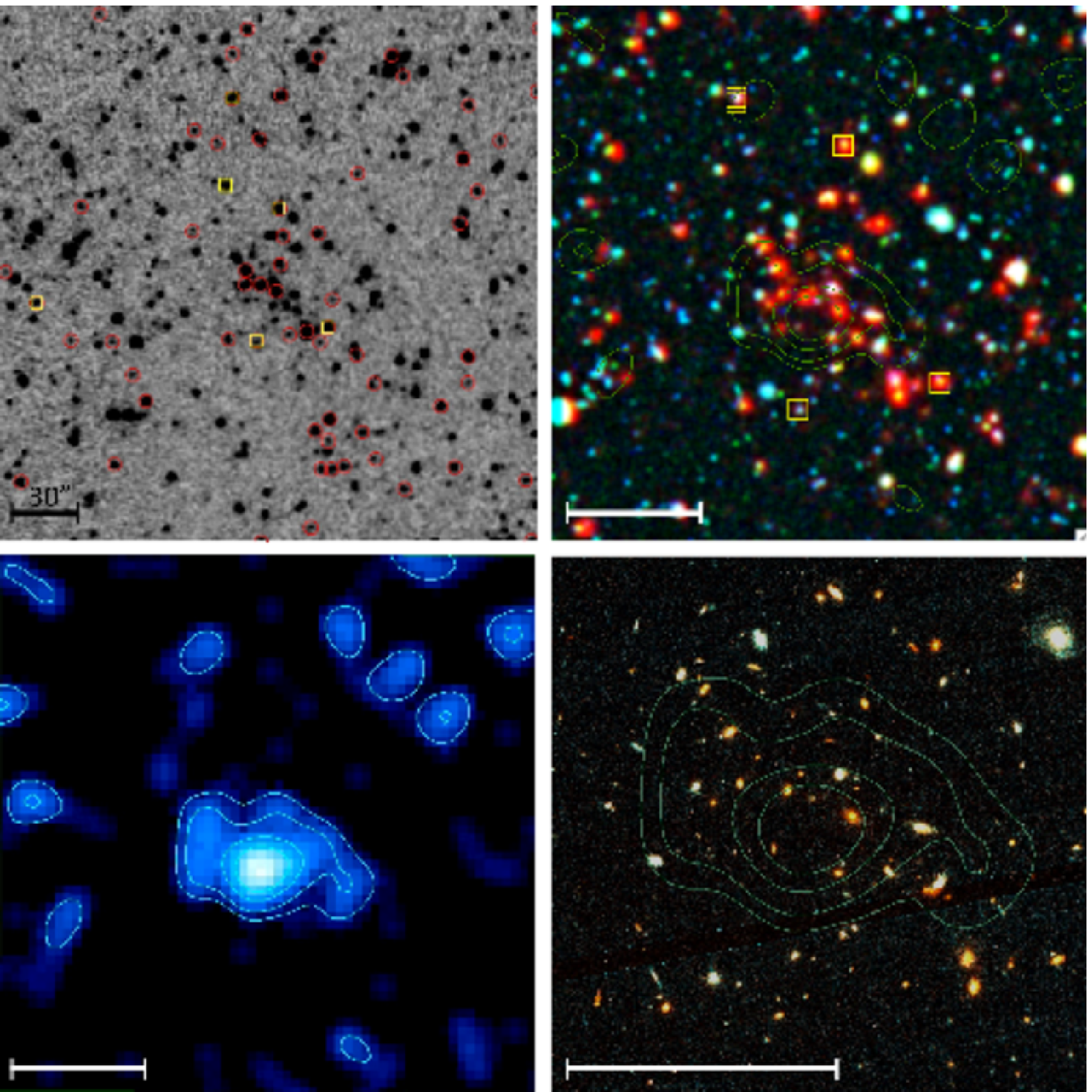}
\caption{Same as for Figure \ref{Fig: image22}, but for cluster \clB\
  at $z=1.487$.  The X-ray contours correspond to 0.06, 0.09, 0.15 and
  0.21 counts per $2\arcsec \times 2\arcsec$ pixel per 31.1 ks in the
  0.7-2 keV band.  The  pseudo-color \hst\ image in the lower right
  panel is composed of ACS/F814W and WFC3/F160W images.
  \label{Fig: image36}}
\end{figure*}

Cluster \clB\ at $z=1.487$ is shown in Figure \ref{Fig: image36} and
Table \ref{Tab: spectra} lists the 8 spectroscopic members secured to
date.  The LRIS spectra were acquired on UT 2010 May 11 using the
G400/8500 grating on the red side, the D680 dichroic, and the 400/3400
grism on the blue side.  The seeing was 0\farcs8 and conditions were
clear.  The WFC3 spectra were obtained on UT 2010 Feb 9 with the G141
grism.  Although the redshift histogram, shown in Figure \ref{Fig:
  zhist}, is currently too sparse to allow a meaningful constraint on
the velocity dispersion, it illustrates that \clB\ has dynamical
properties similar to \clA.

\begin{deluxetable*}{lcclllrcl}
  \tablecolumns{9}
  \tabletypesize{\small} \tablecaption{Spectroscopic Cluster Members\label{Tab: spectra}}
  \tablewidth{0pt} \tablehead{ \colhead{} & \colhead{R.A.} &
    \colhead{Decl.} & \colhead{} &
    \colhead{} & \colhead{} & \colhead{} & \colhead{Exposure Time} & \colhead{}\\
    \colhead{ID} & \colhead{(J2000)} & \colhead{(J2000)} &
    \colhead{z} &    \colhead{$\Delta$z} &
    \colhead{Instrument} &
    \colhead{UT Date} & \colhead{(s)} & \colhead{Reference}
  } \startdata
\cutinhead{\clA; $\left<z\right> = 1.414$}
 J143803.7+341328  &  14:38:03.72  & 34:13:28.2  & 1.410  & 0.009 & \hst/WFC3   & 22 May 2010  & 2011  & 1,2 \\
 J143807.2+341446  &  14:38:07.26  & 34:14:46.3  & 1.410  & 0.009 & \hst/WFC3   & 22 May 2010  & 2011  & 1,2 \\
 J143807.9+341330  &  14:38:07.90  & 34:13:30.7  & 1.426  & 0.009 & \hst/WFC3   & 22 May 2010  & 2011  & 1,2 \\
 J143808.1+341453  &  14:38:08.18  & 34:14:53.6  & 1.398  & 0.009 & \hst/WFC3   & 22 May 2010  & 2011  & 1,2 \\
 J143808.3+341415  &  14:38:08.30  & 34:14:15.1  & 1.420  & 0.009 & \hst/WFC3   & 22 May 2010  & 2011  & 1,2 \\
 J143809.8+341345  &  14:38:09.83  & 34:13:45.0  & 1.423  & 0.009 & \hst/WFC3   & 22 May 2010  & 2011  & 1,2 \\ 
 J143810.0+341421  &  14:38:10.08  & 34:14:21.8  & 1.401  & 0.009 & \hst/WFC3   & 22 May 2010  & 2011  & 1,2 \\
 J143808.3+341417  &  14:38:08.36  & 34:14:17.8  & 1.418  & 0.001 & Keck/DEIMOS & 5 April 2007 & 12600 & 3,4,5 \\ 
 J143810.6+341247  &  14:38:10.64  & 34:12:47.3  & 1.411  & 0.001 & Keck/DEIMOS & 5 April 2007 & 12600 & 3,4,5  \\ 
 J143815.7+341659  &  14:38:15.79  & 34:16:59.4  & 1.412  & 0.001 & Keck/DEIMOS & 5 April 2007 & 12600 & 3 \\ 
 J143816.4+341625  &  14:38:16.41  & 34:16:25.7  & 1.414  & 0.001 & Keck/DEIMOS & 5 April 2007 & 12600 & 3 \\ 
 J143816.8+341440  &  14:38:16.84  & 34:14:40.4  & 1.412  & 0.001 & Keck/DEIMOS & 5 April 2007 & 12600 & 3 \\ 
 J143806.9+341433  &  14:38:06.97  & 34:14:33.8  & 1.415  & 0.001 & Keck/LRIS   & 4 June 2005  & 12600 & 6 \\ 
 J143807.8+341441  &  14:38:07.80  & 34:14:41.6  & 1.417  & 0.001 & Keck/LRIS   & 4 June 2005  & 12600 & 6 \\ 
 J143811.2+341256  &  14:38:11.24  & 34:12:56.2  & 1.417  & 0.001 & Keck/LRIS   & 4 June 2005  & 12600 & 6 \\
 J143812.0+341318  &  14:38:12.09  & 34:13:18.2  & 1.415  & 0.001 & Keck/LRIS   & 4 June 2005  & 12600 & 6 \\
 J143813.3+341452  &  14:38:13.36  & 34:14:52.9  & 1.403  & 0.001 & Keck/LRIS   & 4 June 2005  & 12600 & 6 \\
\cutinhead{\clB; $\left<z\right> = 1.487$}                                                     
J143225.5+325042  &  14:32:25.59  & 32:50:42.3  & 1.499 & 0.009 & \hst/WFC3 &  9 Feb 2010 & 2011 & 1,2 \\ 
J143221.9+324939  &  14:32:21.96  & 32:49:39.5  & 1.479 & 0.001 & Keck/LRIS & 11 May 2010 & 3600 & 1,7 \\ 
J143232.2+324950  &  14:32:32.27  & 32:49:50.2  & 1.485 & 0.001 & Keck/LRIS & 11 May 2010 & 3600 & 1,7 \\ 
J143222.6+325221  &  14:32:22.60  & 32:52:21.6  & 1.483 & 0.001 & Keck/LRIS & 11 May 2010 & 5400 & 1,7 \\ 
J143223.6+325032  &  14:32:23.69  & 32:50:32.7  & 1.479 & 0.001 & Keck/LRIS & 11 May 2010 & 5400 & 1,7 \\ 
J143224.4+324933  &  14:32:24.48  & 32:49:33.6  & 1.486 & 0.001 & Keck/LRIS & 11 May 2010 & 5400 & 1,7 \\ 
J143225.3+325121  &  14:32:25.38  & 32:51:21.9  & 1.496 & 0.001 & Keck/LRIS & 11 May 2010 & 5400 & 1,7 \\
J143225.6+325043  &  14:32:25.60  & 32:50:43.3  & 1.493 & 0.001 & Keck/LRIS & 11 May 2010 & 5400 & 1,7 \\ 
\enddata
\tablerefs{1 (This Work); 2 (Stanford \etal\ in prep); 3  \citep{eisenhardt08}; 4  \citep{dawson09}; 5  \citep{barbary10}; 6  \citep{stanford05}; 7  (Brodwin \etal\ in prep).}
\end{deluxetable*}

\section{X-ray Observations}
\label{Sec: x-ray}

A 145.0~ks observation of \clA\ was obtained in October 2009 by
S.~Andreon using the Advanced CCD Imaging Spectrometer (ACIS-S) on
board the {\it Chandra X-Ray Observatory}
\citep{andreon_chandra_prop08}.  Those data were split between an
observation of 101.3~ks on UT 2009 October 04 (OBS-ID~10461) and an
observation of 43.7~ks on UT 2009 October 09 (OBS-ID~12003).  We
obtained a 35.0~ks ACIS-I observation of \clB\ on UT 2008 October 30
(OBS-ID~10457; PI Stanford).  For the current analysis we do not
include the shallow 5~ks observations of these fields obtained by the
XBo\"otes survey \citep{murray05} for which our faint targets are
off-axis.

We processed the data following standard procedures using the {\it
  Chandra} Interactive Analysis of Observations (CIAO; V4.2) software.
We initially identified the good time intervals for the exposures,
yielding effective exposure times of 144.2~ks for \clA\ and 31.1~ks
for \clB.  Both high redshift clusters are clearly associated with
extended X-ray emission in the reduced images.  After first masking
point sources, we extracted source counts in the 0.6 -- 6 keV range
within 1\arcmin\ radius apertures of each cluster.  This extraction
aperture approximately corresponds to 500~kpc at the cluster
redshifts.  Response matrices and effective areas were then determined
for each detected source.  We used XSPEC (V12.6.0) to fit the
background-subtracted X-ray spectra with the MEKAL hot, diffuse gas
model \citep{mewe85} using the Wisconsin photo-electric absorption
cross-section \citep{morrison83}.  Metal abundances of 0.3 solar were
assumed for both clusters.

Extended X-ray emission was also detected from two additional
structures in the deep \chandra\ exposure.  These have been identified
within the ISCS cluster sample as groups at $z=0.65$ and $z=0.54$, and
will be further discussed by Miller \etal\ (in prep).

\subsection{\clA}

We measure 312 background-subtracted source counts for cluster \clA,
corresponding to a detection significance of 9.7$\sigma$.  The X-ray
peak is centered at $\alpha = 14^h38^m08.3^s$, $\delta = 34\arcdeg
14\arcmin 15\arcsec$ (J2000). Adopting a Galactic $N_H = 9.94 \times
10^{19}~ {\rm cm}^{-2}$ and redshift $z = 1.41$ for \clA, we derive a
gas temperature of $3.3^{+1.9}_{-1.0}$~keV and a Galactic absorption
corrected soft ($0.5 - 2$~keV) X-ray flux of $S_{0.5 - 2} =
9.0^{+2.1}_{-1.7} \times 10^{-15}~ {\rm erg}~ {\rm cm}^{-2}~ {\rm
  s}^{-1}$.  These errors represent the 68\% confidence interval.
\citet{Markevitch98} report that temperature biases that arise in
cool-core clusters can be minimized by excluding the central 70 kpc
from the fit.  We have verified that the excising the core has no
significant effect on the measured temperature in \clA.  The
corresponding rest-frame soft X-ray luminosity for our adopted
cosmology is $L_{0.5 - 2} = 1.00^{+0.24}_{-0.19} \times 10^{44}~ {\rm
  erg}~ {\rm s}^{-1}$.

Adopting the $\Mfive-L_X$ relation of \citet{vikhlinin09}, we derive a
luminosity-based total mass of $\logmlxfive =
14.17\,^{+0.11}_{-0.14}$.  To estimate a total virial mass, we assume
an NFW density profile \citep{nfw} and the mass-concentration relation
of \citet{duffy08}, finding $M_{\mbox{\small v}} \approx $ \Mtwolx $ =
1.54\, $\Mfivelx.  Therefore, $\logmlxtwo = 14.35\,^{+0.11}_{-0.14}$.
We also infer the mass from the $\Mfive-T_X$ relation of
\citet{vikhlinin09}, which has a lower intrinsic scatter than the
luminosity--mass relation and hence provides a useful consistency
check.  The $T_X$-based mass is $\logmtxtwo = 14.2\,^{+0.3}_{-0.4}$.

\subsection{\clB}

We measure 42 background-subtracted source counts for cluster \clB,
corresponding to a detection significance of 5.2$\sigma$.  The X-ray
peak is centered at $\alpha = 14^h32^m24.2^s$, $\delta = 32\arcdeg
49\arcmin 55\arcsec$ (J2000). Adopting a Galactic $N_H = 8.71 \times
10^{19}~ {\rm cm}^{-2}$ and redshift $z = 1.49$ for \clA, we derive a
poorly constrained gas temperature $T_X \sim ~ 2.6$~keV and a Galactic
absorption corrected soft X-ray flux of $S_{0.5 - 2} =
1.0^{+0.8}_{-0.5} \times 10^{-14}~ {\rm erg}~ {\rm cm}^{-2}~ {\rm
  s}^{-1}$. The corresponding rest-frame soft X-ray luminosity is
$L_{0.5 - 2} = 1.3^{+1.1}_{-0.6} \times 10^{44}~ {\rm erg}~ {\rm
  s}^{-1}$, from which we infer a total mass of $\logmlxtwo = 14.4 \pm
0.2$.
 

\section{Discussion}
\label{Sec: discussion}

\subsection{Comparison of Mass Measures}
\label{Sec: mass comparison}

In the preceding sections we presented dynamical and X-ray measures of
the total masses of clusters \clA\ and \clB.  The large number of
spectroscopic redshifts and deep X-ray imaging available for \clA\
allows a comparison of these masses, and therefore, a quantitative
characterization of the ISCS cluster selection at $z\sim 1.4$.  As the
data for \clB\ are not yet of the same quality, we make a more
qualitative mass comparison for this cluster.

A summary of the measured masses is presented in Table \ref{Tab:
  masses}.  The luminosity-based X-ray mass for \clA\ is in excellent
agreement with both the dynamical and $T_X$-based masses.  This
suggests that the $L_X$-based mass, $\logmlxtwo =
14.35\,^{+0.11}_{-0.14}$, is unbiased and we take it as our final mass
estimate for \clA.  While the sparse redshift sampling of \clB\
precludes a direct dynamical mass measurement, the velocity
distribution is similar to that of \clA.  The $L_X$-based mass,
$\logmlxtwo = 14.4 \pm 0.2$, is consistent with the qualitative
expectation from these velocities and confirms that \clB\ is a
similarly massive cluster.  

\begin{deluxetable}{lcc}
  \tabletypesize{\small} \tablecaption{Comparison of Dynamical and X-ray Masses\label{Tab: masses}}
  \tablewidth{0pt} \tablehead{ \colhead{Property} &
    \colhead{\clA} & \colhead{\clB}  } \startdata
$z$                           & $1.414$                & $1.487$\\
$\sigma$ (km s$^{-1}$)        & $757\,^{+247}_{-208}$  & - \\
$L_X$ ($10^{44}$ erg s$^{-1}$) & $1.00^{+0.24}_{-0.19}$  & $1.3^{+1.1}_{-0.6}$\\
$T_X$ (keV)                   & $3.3^{+1.9}_{-1.0}$ & - \\ 
$\logmdyntwo$                 & $14.4\,^{+0.3}_{-0.7}$  &  - \\ 
$\logmlxtwo$                  & $\mathbf{14.35\,^{+0.11}_{-0.14}}$ & $\mathbf{14.4 \pm 0.2}$\\
$\logmtxtwo$                  & $14.2\,^{+0.3}_{-0.4}$ &  - 
\\
\enddata
\end{deluxetable}

\subsection{$\sigma - T_X$ Relation}
\label{Sec: sigma-tx}

There is a well-known correlation between X-ray temperature and galaxy
velocity dispersion for galaxy clusters \citep{lubin93, bird95,
  girardi96, horner99, xue00, ortiz-gil04}.  Although this relation
was defined at low redshift, there is no expectation of, nor evidence
for, evolution in this relation to $z\sim1$
\citep[\eg][]{wu98,tran99,brodwin10}. 

\begin{figure}[bthp]
\epsscale{1.2}
\plotone{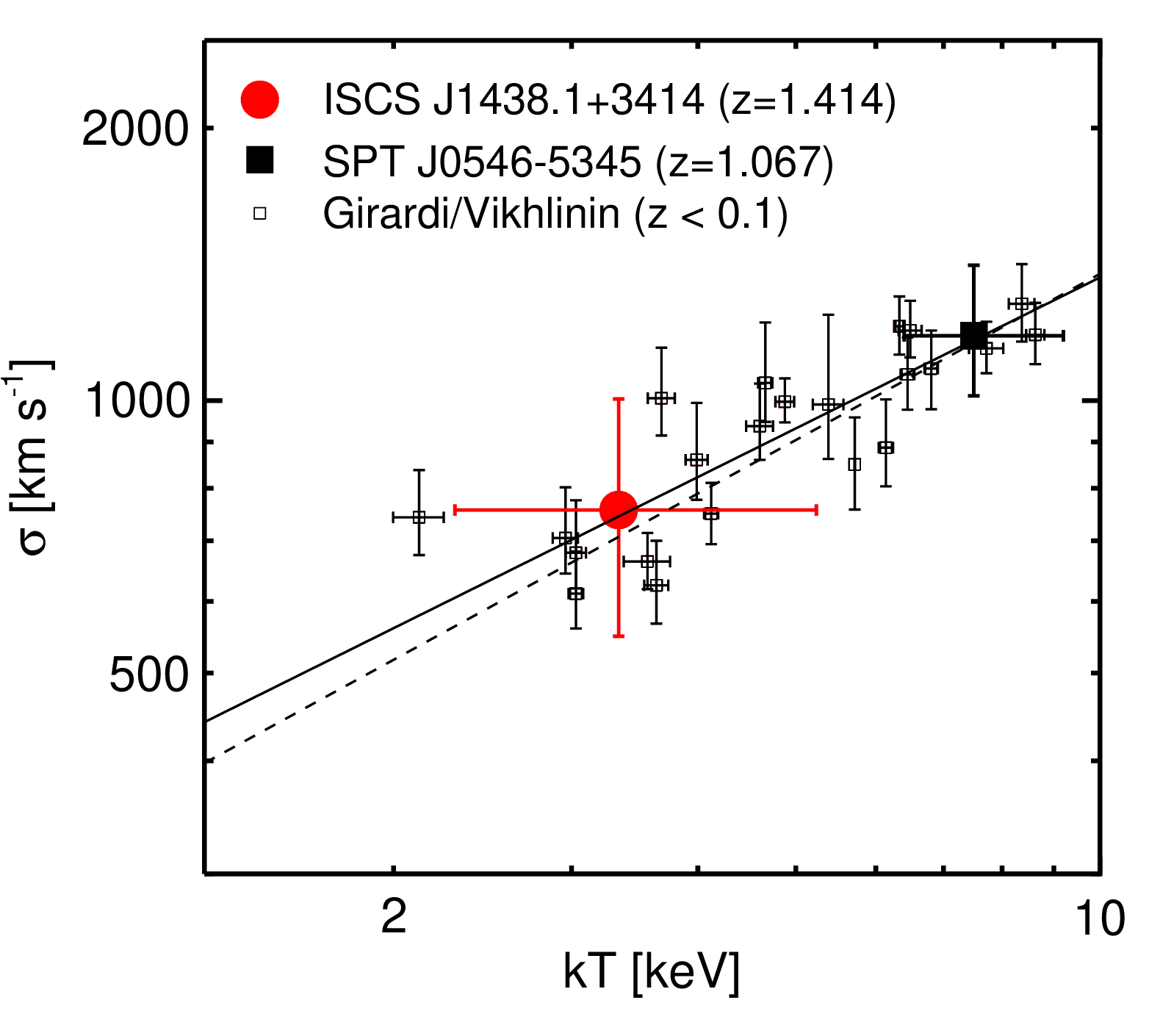}
\caption{Empirical $\sigma - T_X$ correlation for the X-ray clusters
  from \citet{vikhlinin09}, with velocity dispersions taken from
  \citet{girardi96}. The filled black square shows the result for
  \clname, a massive SZE-selected cluster at $z=1.067$
  \citep{brodwin10}.  The fits, described in the text, are typical of
  those found by other authors \citep[\eg][]{lubin93,horner99}. \clA\
  (red circle) is clearly consistent with this relation.}
  \label{Fig: sigma-Tx}
\end{figure}
 
Figure \ref{Fig: sigma-Tx} plots \clA\ on this $\sigma-T_X$ relation
(large red circle).  At $z=1.414$, this is the most distant cluster
for which this relation has ever been tested.  A comparison sample of
low redshift ($z<0.1$) clusters is plotted, along with the most
massive spectroscopically confirmed cluster at $z>1$, \clname\ from
\citet{brodwin10}. The dashed line is the best-fit relation
\citet{girardi96} obtained using previous temperatures from the
literature, and the solid line is our own fit using the
\citet{vikhlinin09} temperatures for the low-z clusters. \clA\ is
clearly consistent with this relation.  We conclude that it is similar
in nature to massive clusters selected in the X-ray or SZE.  In
particular, it is not underluminous in the X-ray relative to its
dynamical properties, as has been suggested of some optically selected
clusters at lower redshifts \citep{bower94,
  donohue01,gilbank04,popesso07}.

\subsection{Systematic Uncertainties}
\label{Sec: systematic}

Total cluster masses inferred from dynamical or X-ray observables are
subject to various systematic uncertainties.  Here we briefly explore
how these might affect our derived cluster masses.

\subsubsection{Dynamical Mass Systematics}

\noindent {\bf Galaxies as Tracers of the Potential~~} While cluster
galaxies are found in simulations to be unbiased tracers of the
gravitational potential \citep[\eg][and references therein]{evrard08},
observations of low redshift clusters indicate that emission line
galaxies (ELGs) may produce dispersions $\sim 10\%$ higher than those
measured from quiescent galaxies
\citep[\eg][]{girardi96,biviano97,detheije99,sodre89}.  This is
thought to be primarily due to the influence of infalling field
galaxies which have not fully virialized.  Of the spectroscopic
members in \clA, 11/17 were secured via emission features, so this is
a possible source of systematic uncertainty.  However, high star
formation rates are observed in clusters at $z\sim 1.5$
\citep{hilton10,tran10}, including amongst the central, massive
members in the cluster cores.  Indeed, the central galaxies in \clA\
have very high star formation rates, as inferred from \spitzer\
24$\mu$m observations (Brodwin \etal\ in prep).  This suggests that
the simple interpretation of ELGs as tracers of the infalling galaxy
population is likely at least partially incorrect in \clA.  It is
therefore not straightforward to quantify the possible bias in the
dispersion due to the population mix at such high redshift, but it is
likely $< 10\%$.  This is much smaller than the statistical errors and
would not affect any of the conclusions presented here.

\vspace{0.15cm}
\noindent {\bf The Virial Approximation~~} Cluster mass estimates based upon
velocity dispersions are predicated upon the assumption of virial
equilibrium, yet for systems at high redshift and with disturbed X-ray
morphologies like \clA\ it is not {\it a priori} obvious that this
approximation is reasonable. In general, departures from virial
equilibrium tend to bias mass estimates upward due to the inclusion of
infalling structures with bulk motions. We cannot discount the
possibility that for this system the dispersion is impacted by
non-virial motion. The agreement between the dynamical, lensing, and
ICM mass determinations, however, argues that the systematic bias from
this effect is not large.

\vspace{0.15cm}
\noindent {\bf Impact of Radial Sampling~~} The observed galaxy
cluster velocity dispersion is sensitive to the radius over which it
is measured since the line-of-sight dispersion decreases with
radius. The radius $\rtwo$, where the overdensity is 200 times the
critical density, is the optimal radius within which to measure the
dispersion for determination of $\Mtwo$
\citep[\eg][]{katgert96,katgert98,rines06,biviano06}.  In \clA\ our
spectroscopy extends to $\sim 1.5 \rtwo$, where we estimate $\rtwo =
1.05 \,^{+0.27}_{-0.20}$ Mpc following \citet{carlberg97}.  According
to the simulations of \citet{biviano06}, this may produce an
underestimate of the dispersion by $\sim 6\%$.  This is both small and
in the opposite sense as the possible bias due to the population mix,
and therefore should not materially affect our results.

\subsubsection{X-ray Mass Systematics}

\noindent {\bf Evolution of Scaling Relations~~} The primary systematic
uncertainty in deriving total masses from the X-ray observables of
temperature and luminosity arises from possible evolution in the
temperature- and luminosity-mass scaling relations.  These relations
have been calibrated primarily at moderate ($z \la 0.6$) redshift
\citep{maughan07, vikhlinin09, mantz10}, and possible evolutionary
effects at $z>1$ are untested.  We therefore simply note that \clA\
and \clB\ provide valuable additions to the handful of
spectroscopically confirmed $z>1.4$ clusters with X-ray detections
\citep[\eg][]{stanford06, papovich10, tanaka10, gobat10}.

\vspace{0.15cm}
\noindent {\bf Cluster Dynamical State~~} Another possible systematic
issue arises from the dynamical state of these high redshift clusters.
As Figures \ref{Fig: image22} and \ref{Fig: image36} illustrate, the
X-ray morphologies of \clA\ and \clB\ are unrelaxed, showing the
filamentary structure and isophotal centroid shifts
\citep[\eg][]{mohr93} typical of most high redshift clusters
\citep{vikhlinin09}.  Using high-resolution numerical simulations,
\citet{kravtsov06} conclude that the $\Mfive-T_X$ relation may
underestimate cluster mass by $\sim 20\%$ in such unrelaxed clusters,
and therefore the temperature-based mass for \clA\ may be
underestimated at this level. Given the considerable uncertainties in
these scaling relations at $z > 1$ we opt not to apply this
correction, but simply note that it would bring the $T_X$-based mass
into even closer agreement with the other mass measures.

\section{Conclusions}
\label{Sec: conclusions}

We present X-ray observations of two $z>1.4$ spectroscopically
confirmed galaxy clusters from the ISCS.  \clA, at $z=1.414$, is
detected in deep \chandra\ data at a significance of $9.7\sigma$.  We
measure a luminosity of $L_{0.5 - 2} = 1.00^{+0.24}_{-0.19} \times
10^{44}~ {\rm erg}~ {\rm s}^{-1}$ and a temperature of $T_x =
3.34^{+1.90}_{-1.04}$~keV.  From these we derive total $L_X$- and
$T_X$-based masses of $\logmlxtwo = 14.35\,^{+0.11}_{-0.14}$ and
$\logmtxtwo = 14.2\,^{+0.3}_{-0.4}$.  These agree with the dynamical
mass, inferred from 17 spectroscopically confirmed members, of
$\logmdyntwo = 14.4\,^{+0.3}_{-0.7}$.  The X-ray and dynamical
properties of \clA\ are fully consistent with the $\sigma-T_X$
relation.  This is the most distant cluster for which this relation
has been tested.

We also present spectroscopic confirmation for \clB\ at $z=1.487$, the
most distant of 18 $z>1$ clusters confirmed in the ISCS to date.
\clB\ is detected at a $5.2\sigma$ significance in a shallow \chandra\
exposure.  We measure an X-ray luminosity of $L_{0.5 - 2} =
1.3^{+1.1}_{-0.6} \times 10^{44}~ {\rm erg}~ {\rm s}^{-1}$, from which
we infer a total mass of $\logmlxtwo = 14.4 \pm 0.2$.  The sparse
dynamical data are consistent with this X-ray mass, confirming that
\clB\ is also a massive high redshift cluster.

Our primary conclusion is that the $z \ga 1.4$ clusters studied in
this work, identified by the stellar-mass selection of the ISCS, are
massive and have dynamical and ICM properties typical of clusters
selected by the X-ray or SZE.  This further confirms that stellar-mass
cluster selection provides a powerful and sensitive method for
studying cluster evolution to the highest redshifts.  The two clusters
presented in this work, with total masses of $\logmtwo \approx 14.4$
or $\approx 2-3 \times 10^{14}$ \msun, are nearly Coma--mass
progenitors.  The ISCS is therefore identifying the precursor
population of present-day massive clusters.

\acknowledgments This work is based in part on observations obtained
with the {\it Chandra X-ray Observatory} (CXO), under contract
SV4-74018, A31 with the Smithsonian Astrophysical Observatory which
operates the CXO for NASA.  Support for this research was provided by
NASA grant G09-0150A.  This work is based in part on observations made
with the {\it Spitzer Space Telescope}, which is operated by the Jet
Propulsion Laboratory, California Institute of Technology under a
contract with NASA. Support for this work was provided by NASA through
an award issued by JPL/Caltech. This work is based, in part, on
observations made with the NASA/ESA {\it Hubble Space Telescope},
obtained at the Space Telescope Science Institute, which is operated
by the Association of Universities for Research in Astronomy, Inc.,
under NASA contract NAS 5-26555. These observations are associated
with programs 11597 and 11663.  Support for programs 11597 and 11663
were provided by NASA through a grant from the Space Telescope Science
Institute, which is operated by the Association of Universities for
Research in Astronomy, Inc., under NASA contract NAS 5-26555.  Some of
the data presented herein were obtained at the W.~M.~Keck Observatory,
which is operated as a scientific partnership among the California
Institute of Technology, the University of California and the National
Aeronautics and Space Administration.  The Observatory was made
possible by the generous financial support of the W.~M.~Keck
Foundation.  This work makes use of image data from the NOAO Deep
Wide--Field Survey (NDWFS) and the Deep Lens Survey (DLS) as
distributed by the NOAO Science Archive. NOAO is operated by the
Association of Universities for Research in Astronomy (AURA), Inc.,
under a cooperative agreement with the National Science Foundation.

This paper would not have been possible without the efforts of the
\spitzer, \chandra, \hst\ and Keck support staff.  Support for MB was
provided by the W.~M.~Keck Foundation.  The work by SAS at LLNL was
performed under the auspices of the U.~S.~Department of Energy under
Contract No. W-7405-ENG-48.


\end{document}